\documentstyle[twocolumn,aps]{revtex}
\begin{document}
\draft
\preprint{Dortmund, Mai 1995}

\title{
The transition from an ordered antiferromagnet
to a quantum disordered spin liquid \\
in a solvable bilayer model
       }
\author{Claudius Gros and Wolfgang Wenzel
       }
\address{Institut f\"ur Physik, Universit\"at Dortmund,
         44221 Dortmund, Germany
        }
\author{Johannes Richter
       }
\address{Institut f\"ur Theoretische Physik,
         Otto-von-Guericke Universit\"at Magdeburg,
         Postfach 4120,
         39016 Magdeburg, Germany.
        }
\date{\today}
\maketitle
\begin{abstract}
We present a spin-1/2 bilayer model for the
quantum order-disorder transition which (i) can be solved
by mean-field theory for bulk quantities, (ii) becomes
critical at the transition, and (iii) allows to include intralayer
frustration. We present numerical data (for systems with up
to 240 sites) and analytical results for the critical coupling
strength, ground-state energy, order parameter and
for the gap. We show that the critical coupling decreases
linearly with frustration.

\end{abstract}
%
%
%
The model of two coupled antiferromagnetic layers, the
`bilayer model', has attracted much attention
in the last years
\cite{Sandvik_Scalapino,Hida,Elstner_95}.
The physics of this model is dominated by the competition
between the in-plane coupling $J_{||}$ and the inter-plane
coupling $J_{12}$. A transition occurs from an ordered
antiferromagnetic state to a spin-liquid state with a gap
at a certain critical ratio $J_c$ of $J_{12}/J_{||}$.
Millis and Monien \cite{Millis_Monien} proposed that
certain anomalies in the magnetic response of $YBa_2Cu_3O_{6+x}$,
interpreted commonly as evidence for a spin gap might be explained
within the bilayer model.

Solvable model Hamiltonians play an important role in condensed matter
theory, as they provide new insights and because they can be used as
references for approximate theories. Here we propose a non-trivial
`long-range' version of the bilayer model for which we are able to
solve for all quantities of interest, while retaining the crucial
ingredients driving the quantum order-disorder transition.  In this
model the in-plane coupling $J_1$ is of long-range Lieb-Mattis type
while the inter-plane coupling, $J_{12}$, remains short ranged.  As a
result, bulk expectation values for this system are accessible by
analytical considerations, we are able to determine the exact
ground-state wavfunction up to intensive corrections. Excitations,
remain non-trivial however, but the simplicity of the in-plane
interaction facilitates the numerical analysis to obtain excellent
control over finite size corrections. We determine the functional
dependence of the gap as a function of $J_{12}/J_1$, presenting the
first rigorous study of the impact of in-plane frustration on the
stability of the spin-liquid phase. We find that the critical $J_c$ is
reduced {\it linearly} with the in-plane frustration
\cite{note_Monien,inui}. This result is of particular interest since a
substantial interlayer coupling constant $J_{12}$ was found in recent
infrared absorption experiments on Y$_1$Ba$_2$Cu$_3$O$_6$
\cite{thilo} and by a detailed quantum chemical calculation
\cite{barri,note_xx}.

{\em Model:} We define the `long-range' bilayer Hamiltonian as
$H_{LR}=\sum_{\gamma}H_{\gamma} +H_{12}$,
with
\begin{equation}
\begin{array}{rcl}
H_{\gamma} &=& (J_1/N)\,
             {\bf S}_{A\gamma}\cdot{\bf S}_{B\gamma}
 + (J_2/N)\, [ {\bf S}_{A\gamma}^2+{\bf S}_{B\gamma}^2]
\\ [2pt]
H_{12} &=&  J_{12} \sum_{i}
            {\bf S}_{i,1}\cdot{\bf S}_{i,2},
\end{array}
\label{H_LR}
\end{equation}
and $J_1,J_2,J_{12}>0$.  Here ${\bf S}_{i,\gamma}$ are the spin-1/2
operators of the first $(\gamma=1)$ and of the second $(\gamma=2)$
layer respectively and ${\bf S}_{A\gamma}=\sum_{i\in A}{\bf
S}_{i,\gamma}$, ${\bf S}_{B\gamma}=\sum_{i\in B}{\bf S}_{i,\gamma}$
are the total-spin operators of the A/B sublattice in the respective
layers. Each sublattice contains N spins, the total number of sites is
$N_s=4N$.

The first term of the intra-layer interaction ($J_1$), favours an
anti-alignment of the sublattice spins, i.e.  an antiferromagnetically
ordered state. The second, frustrating, intra-layer interaction
($J_2$) favours minimal sublattice magnetisation and therefore tends
to suspress the antiferromagnetic order.  The local inter-plane
coupling term $H_{12}$ favours the formation of local singlets. As a
result, this model contains the relevant interactions to exhibit a
quantum order-disorder transition as the relative strength of intra-
and inter-layer interactions are varied. For small values of the
interlayer coupling $J_{12}$ the ground-state is characterized by
long-range antiferromagnetic order and gapless magnetic excitations.
The spin-liquid state, realized for large $J_{12}$, is characterized
by a finite spin gap and the absence of antiferromagnetic order.

The long-range bilayer Hamiltonian $H_{LR}$
has a close relation to its `short-range' counter-part,
$H_{SR}=\sum_\gamma H_\gamma^{SR}+H_{12}$, with
\begin{equation}
H_\gamma^{SR}  =
  J_{||} \sum_{<i,j>}
   {\bf S}_{i,\gamma}\cdot {\bf S}_{j,\gamma}
  +  J_{||}^{\prime}
 \sum_{[i,l]}
   {\bf S}_{i,\gamma}\cdot{\bf S}_{l,\gamma}
\label{H_SR}
\end{equation}
where the symbols $<i,j>$ and $[i,l]$ denote pairs of n.n. and n.n.n
sites on a square lattice respectively and where the terms $\sim
J_{||}$ and $\sim J_{||}^\prime$ correspond to the term $\sim J_1$ and
$\sim J_2$ in (\ref{H_LR}) respectively.  In order to make connection
between $H_{LR}$ and $H_{SR}$ we compare the energies of the N\'eel
state in both models for the case $J_2=J_{||}^{\prime}=0$. This
introduces a correspondence relation $J_1\hat{=} 4 J_{||}$ between the
respective couplings. In $H_{SR}$ the order-disorder transition occurs
at $J_{12}/J_{||}\approx 2.51$ \cite{Sandvik_Scalapino}.  We expect
the true critical interlayer coupling for $H_{LR}$ to be somewhat
larger than the value $J_{12}/J_1\approx 2.51/4\approx 0.628$
suggested by this naive rescaling, since intra-plane quantum
fluctuations are suspressed by the long-range nature of $J_1$.

{\em Results:}
In the limit of decoupled layers $J_{12}=0$ the expectation
value of the energy per layer is
\[
{J_1\over 2N}\,S_1(S_1+1)
+ {2J_2-J_1\over2N}[ S_{A1}(S_{A1}+1)
           + S_{B1}(S_{B1}+1)].
\]
where $S_{A1},S_{B1}$ are the eigenvalues of ${\bf S}_{A1}$, ${\bf
S}_{B1}$ and of the total-spin operator ${\bf S}_{A1}+{\bf S}_{B1}$
respectively \cite{LM,richter}. For $J_2<J_1/2$ the ground state is a
singlet with $S_1=0$ and maximal sublattice magnetization,
$S_{A1}=S_{B1}=N/2$. In the thermodynamic limit the ground-state
energy per site is then $(2J_2-J_1)/8$.  In the opposite limit $J_{12}
= \infty$ the ground state is a product of interlayer singlet pairs
with energy $-3/8J_2$ per site and a gap of order $J_{12}$ to the
lowest excited state, a spin triplet.
%
%
%

{\em Variational Analysis:} In the presence of a finite interlayer
coupling a natural ansatz for a variational wavefunction interpolates
between the two extremal regimes:
\begin{equation}
\label{psi}
|\Psi(\alpha)>\ =\
\prod_{i}(1-\alpha [S_{i,1}^-S_{i,2}^+
                   +S_{i,1}^+S_{i,2}^-])\,|AF>,
\end{equation}
where $|AF>$ is the N\'eel state \cite{Neel}.  A straightforward
minimization of the energy yields the critical coupling constant
$J_c=J_1-2J_2$. The optimal value for the variational parameter is
found to be $\alpha\equiv1$, for $J_{12}>J_c$, and to be
$\alpha=J_c/J_{12}-\sqrt{(J_c/J_{12})^2-1}$ for $J_{12}<J_c$.

Of interest are the expectation values of the ground-state energy per site:
\begin{equation}
e_N = <H>_0/(4N)+3/8J_{12},
\label{e}
\end{equation}
where we have subtracted the energy of the
product state realized for $J_{12}\rightarrow\infty$,
and for the in-plane staggered order parameter,
\begin{equation}
o_N\ =\ \sqrt{
             |<{\bf S}_{A1}\cdot{\bf S}_{B1}
           + {\bf S}_{A2}\cdot{\bf S}_{B2}>|}\,/\,(4N).
\label{o}
\end{equation}
The variational ground-state energy per site (\ref{e}) and the order
parameter (\ref{o}) are then computed to be
\begin{eqnarray}
\label{e_0}
e_0 &=& {-J_c\over8}[1-J_{12}/J_c]^2 \\
\label{o_0}
o_0 &=& {1\over\sqrt{32}}\sqrt{1-(J_{12}/J_c)^2}
\end{eqnarray}
for $J_{12}<J_c$. For $J_{12}>J_c$ both $e_0$ and $o_0$ are
zero. For $J_{12}>J_c$ the ground state is a product of inter-layer
spin singlets, for $J_{12}<J_c$ an ordered antiferromagnet.

The quality of a trial wavefunction $|\Psi>$ is measured by its variance,
\begin{equation}
\sigma\ =\ {<\Psi|(H-<H>)^2|\Psi> \over <\Psi|\Psi>},
\label{sigma}
\end{equation}
which is zero for an exact eigenstate of $H$. In general, variational
wavefunctions which are not exact eigenstates lead to an {\it
extensive} variance, $\sigma\sim O(N)$ \cite{criterion}.  For the
wavefunction given by (\ref{psi}), however, we find an intensive
variance, $\sigma(\alpha)\sim O(N^0)$, which indicates that the
expressions for the
variational ground-state energy $e_0$, Eq.\ \ref{e}, and order
parameter, Eq.\ \ref{o}, are {\em exact} in the thermodynamic limit.
However, since the energy is known only up intensive correction, the
gap cannot be determined in this framework.

{\em Numerical Analysis:} The Hamiltonian (\ref{H_LR}) is separately
invariant under any permutation within the A-sublattice and within the
B-sublattice. The ground-state singlet and the lowest excited triplet
state belong to the symmetric subspace of the permutation symmetry, as
we found by diagonalizing small clusters with $N_s=4 - 20$ spins
within the full Hilbert space.  We conjecture that the
symmetry of the lowest singlet state and of the lowest triplet state
does not change for $N_s > 20$. Within the symmetric representation the
size of the Hilbert space grows only $\sim N_s^5$, i.e.  algebraically.
We have numerically diagonalized systems with
up to $N_s=4N=240$ sites, using a specially adapted  exact diagonalization
technique \cite{wenzel}.  The results for the ground-state energy per
site,
are presented in Fig.\ \ref{e_o} for $J_2/J_1=0,\ 0.2$. The width of
the transition from the ordered state at small $J_{12}/J_1$ to the
quantum disordered spin liquid at large $J_{12}/J_1$ is attributable
to finite size effects \cite{Kittel_Shore,Botet_et_al}.

Using finite-size scaling we have estimated the critical exponents of
the finite-size corrections to the data for the order parameter
presented in Fig.\ \ref{e_o}. We found the finite-size corrections to
scale like $N^{-1}$, $N^{-1/3}$ and $N^{-1/2}$ for $J_{12}<J_1$,
$J_{12}=J_1$ and $J_{12}>J_1$ (with $J_2=0$) respectively.  A similar
behaviour has been found for the long range Ising model in a
transversal field \cite{Botet_et_al}.

We have extrapolated the numerical data obtained from clusters with up
to 240 sites to the thermodynamic limit via the formula $\
\lim_{N\to\infty}e_N=e_{\infty}+B/N+C/N^2+\dots$, where we have used
the $1/N$ finite size scaling of the Lieb-Mattis antiferromagnet as
the leading term\cite{LM,richter}. The $1/N$ scaling for the
finite-size corrections of the ground-state energy is also found in a
least square fit to the numerical data presented in Fig.\ \ref{e_o}.
For all parameter values considered, i.e for
$J_{12}/J_1 = 0,\, 0.1,\dots, 1.0$ and for $J_2/J_1 = 0,\,
0.125,\dots 0.5$, we found agreement to at least six digits between
the such obtained $e_{\infty}$ and the analytical formula for $e_o$
(\ref{e_0}). This very good agreement between the numerical and
analytical results gives us confidence that the gap, which is
accessible only to numerical calculation can be determined with
similar precision.

%
%
%
In Fig.\ \ref{gap} we present the numerical results of the gap,
$\Delta_N$, for clusters with up to $N_s=240$ sites and $J_2=0$
together with the mean-field prediction for the gap,
\begin{equation}
\Delta_{MF} \ =\  {J_c+J_{12}\over 2}\,\theta(J_c-J_{12})\ + \
                       J_{12}        \,\theta(J_{12}-J_c).
%
%
%
%
\label{gap_MF}
\end{equation}
Due to the spin-rotational invariance of (\ref{H_LR}) the gap
$\Delta_N$ differs non-trivially from the mean-field result. Note,
that this is possible only because the variance (\ref{sigma}) of the
mean-field solution (\ref{psi}) is intensive and not zero.
Fig.\ \ref{gap} also contains  a conjecture for the gap in the
thermodynamic limit,
\begin{equation}
\Delta_0\ =\ J_{12} \sqrt{1-J_c/J_{12}},
\label{gap_0}
\end{equation}
for $J_{12}>J_c=J_1-2J_2$. In the insert of Fig.\ \ref{gap} we have
plotted $\log(\Delta_N-\Delta_o)$ versus $\log(1/N)$. We note that the
data fall on a straight line with slope one for all $J_{12}>J_c$,
suggesting that (\ref{gap_0}) is indeed the correct expression for the
gap in the thermodynamic limit. At the critical point,
$J_{12}=J_1-2J_2$, the finite-size corrections scale like
$(1/N)^{\nu}$. We estimate numerically $\nu=0.3$, a value which
deviates from mean-field predictions and for which we do not have
an explanation at present.

%
%
%

{\em Discussion:} In (\ref{o}) we defined the order parameter $o_N$ of
the bilayer system. It is of interest to examine with $\tilde o_{N}\
=\ \sqrt{ |<{\bf S}_{A1}\cdot{\bf S}_{B2} +{\bf S}_{B1}\cdot{\bf
S}_{A2}>|}\,/\,(4N)$ the relative ordering of the two layers.  An
inspection of the variational wavefunction (\ref{psi}) shows that
$o_N$ and $\tilde o_N$ are directly proportional in the thermodynamic
limit.  A similar result has been found in a numerical study of the
short range model \cite{Sandvik_Scalapino}.  The equivalence of $o_N$
and $\tilde o_N$ implies that the magnetic order parameters of the two
layers are aligned for any $0<J_{12}<J_c$.  This type of magnetic
correlation has been found in neutron scattering experiments on YBaCuO
system \cite{ybaco}. Physically we understand that any $J_{12}>0$
lifts the degeneracy of the two intra-layer N\'eel states and
correlates the two layers magnetically.

In the long range bilayer model the order-disorder transition is
second order with mean-field exponents for ground-state properties \cite{chub}.
The gap opens continuously on the disordered side, with an exponent
for the finite-size correction which deviates from the predictions
of mean-field theory. While the in-plane quantum
fluctuations are too weak to influence bulk properties like the
ground-state energy, they are important for the gap between the exact
ground-state, which is an overall spin singlet, and the lowest excited
triplet state.  In the asymptotic limit
$J_{12}/J_1\rightarrow\infty$ the gap (\ref{gap_0}) is $J_{12}$
but for any $J_{12}/J_1<\infty$ the gap is reduced by the in-plane
quantum fluctuations and is strictly smaller than $J_{12}$.

Our rigorous result that the critical coupling strength depends
{\em linearly} on the frustration is consistent with a
recent effective-action mean-field approach \cite{dotsenko} to the
frustrated bilayer model \cite{note_Monien}. Furthermore this
results does support, to some extend, the possible explanation of the
observed spin-gap in the YBaCuO System,
in view of the recent estimate
of $J_{12}/J_1\approx 0.55$ by infrared absorption \cite{thilo},
by a doping-frustrated bilayer system \cite{Millis_Monien}.
Let us also note in this context that the linear reduction of the
sublattice magnetization by the inter-layer coupling (\ref{o_0})
is due to quantum fluctuations only and that the order-disorder
transition found in $H_{LR}$ has no classical analagon. Eq. (\ref{o_0})
represents therefore the first rigorous result for the reduction
of a classical order paramter by quantum fluctuations in dimensions
greater than one.

In summary we have presented a novel, solvable model for the quantum
order-disorder transition. The quantity of interest, the spin gap, is
critical at the transition while the bulk properties of this model can
be determined analytically. The intra-layer quantum fluctuations are
only {\em intensive}, which allows a complete characterization of the
qunatum order-disorder transition. The inter-layer quantum
fluctuations remain {\em extensive}, introducing the criticality
neccessary for the transition.


This work was supported by the Deutsche Forschungsgemeinschaft,
by the Minister f\"ur Wissenschaft
und Forschung des Landes Nordrhein-Westfalen and by the
European Community European Strategic Program for
Research in Information Technology program, Project
No. 3041-MESH.
%
%
%

%
%
%
\begin{figure}
\caption{ For $N=10,20,30,40,50,60$ and
          a) $J_2/J_1=0$ and b) $J_2/J_1=0.2$
          the numerical results for the order parameter
          (\protect\ref{o})
          (open squares, smallest $N$ at the top)
          and for the ground-state energy per site
          (\protect\ref{e})
          in units of $J_1$
          (open circles, smallest $N$ at the bottom).
          The dashed lines are the analytic expressions
          corresponding to Eq.\
          (\protect\ref{e_0}) and Eq.\ (\protect\ref{o_0}).
          }
\label{e_o}
\end{figure}
\begin{figure}
\caption{ For $N=10,20,30,40,50,60$ and
          $J_2/J_1=0$ the numerical results for the
          gap $\Delta_N$
          (open circles, smallest $N$ at the top).
          The dashed line is the analytic expression
          corresponding to Eq.\ (\protect\ref{gap_0}),
          the dotted line is the mean-field prediction,
          Eq.\ (\protect\ref{gap_MF}).
          Insert: A log-log of the data
          for $J_{12}/J_1= 1.0,\ 1.2,\ 1.6,\ 2.0$.
          }
\label{gap}
\end{figure}

\begin{references}
\bibitem{Neel} Let $i\in A $ and $j\in B$. The
N\'eel state is then defined by
$S_{i,1}^z|AF>=+1/2|AF>$ and
$S_{j,1}^z|AF>=-1/2|AF>$ for the first layer
and by
$S_{i,2}^z|AF>=-1/2|AF>$ and
$S_{j,2}^z|AF>=+1/2|AF>$ for the second layer.
\end{references}
\end{document}